\documentclass[conference, final]{IEEEtran}
\IEEEoverridecommandlockouts
\usepackage{cite}
\usepackage{amsmath,amssymb,amsfonts}
\usepackage{algorithmic}
\usepackage{graphicx}
\usepackage{textcomp}

\usepackage[inline]{enumitem}
\usepackage{xcolor}

\usepackage{witharrows}

\usepackage[flushleft]{threeparttable}
\usepackage{tablefootnote}
\usepackage{array}

\newcolumntype{P}[1]{>{\centering\arraybackslash}m{#1}}
\usepackage{multicol}
\usepackage{multirow}
\usepackage{tabulary}
\usepackage{colortbl}
\definecolor{mygray}{gray}{0.90}
\usepackage{makecell}
\usepackage{booktabs}
\usepackage{subcaption}
\usepackage{stmaryrd}
\usepackage{float}
\usepackage{pifont}

\usepackage{arydshln}
\colorlet{mygray}{gray!15!white}

\usepackage[switch,columnwise]{lineno}

\usepackage{url,hyperref,microtype}
\hypersetup{
    colorlinks=true,
    citecolor=blue,
    linkcolor=blue,
    filecolor=magenta,      
    urlcolor=black,
}

\def\BibTeX{{\rm B\kern-.05em{\sc i\kern-.025em b}\kern-.08em
    T\kern-.1667em\lower.7ex\hbox{E}\kern-.125emX}}
\begin{document}

\title{Beyond the Raw Waveform: Fusing Visual Representations of EDA for Stress Detection\\
}

\author{

\IEEEauthorblockN{Stefanos Gkikas}
\IEEEauthorblockA{\textit{Honda Research Institute Japan} \\
Wako City, Japan \\
stefanos.gkikas@jp.honda-ri.com}

\and

\IEEEauthorblockN{Thomas Kassiotis}
\IEEEauthorblockA{\textit{Department of Electronic Engineering} \\
\textit{Hellenic Mediterranean University}\\
Chania, Greece \\
ddk305@edu.hmu.gr}

\and

\IEEEauthorblockN{Yang Guo}
\IEEEauthorblockA{\textit{Faculty of Information Science}\\
\textit{and Engineering} \\
\textit{Ocean University of China}\\
Qingdao, China \\
diw85827@gmail.com}

\and

\IEEEauthorblockN{Guangliang Li}
\IEEEauthorblockA{\textit{Faculty of Information Science}\\
\textit{and Engineering} \\
\textit{Ocean University of China}\\
Qingdao, China \\
guangliangli@ouc.edu.cn}

\and

\IEEEauthorblockN{Eric Nichols}
\IEEEauthorblockA{\textit{Honda Research Institute Japan} \\
Wako City, Japan \\
e.nichols@jp.honda-ri.com}

\and

\IEEEauthorblockN{Houshyar Asadi}
\IEEEauthorblockA{\textit{Institute for Intelligent Systems Research} \\
\textit{and Innovations (IISRI)} \\
\textit{Deakin University}\\
Geelong, Australia \\
houshyar.asadi@deakin.edu.au}

\and

\IEEEauthorblockN{Nikolaos Smyrnis}
\IEEEauthorblockA{\textit{University Mental Health, Neurosciences and}\\
\textit{Precision Medicine, Research Institute ``COSTAS STEFANIS''} \\
\textit{2nd Department of Psychiatry, Medical School,} \\ 
\textit{National and Kapodistrian University of Athens} \\
Athens, Greece \\
smyrnis@med.uoa.gr}

\and

\IEEEauthorblockN{Giorgos Giannakakis}
\IEEEauthorblockA{\textit{Department of Electronic Engineering} \\
\textit{Hellenic Mediterranean University}\\
Chania, Greece \\
ggian@hmu.gr}

}

\maketitle

\begin{abstract}
Electrodermal activity (EDA) is widely used in automatic stress detection, yet most pipelines treat it only as a raw one-dimensional waveform. This study examines whether complementary visual representations of EDA provide useful information for stress classification and whether their fusion improves recognition performance.
Six image-based representations are derived from each EDA recording: an unwrapped short-time Fourier transform (STFT) phase spectrogram, an instantaneous-frequency map computed from that phase, a power spectral density (PSD) spectrogram, a continuous wavelet transform scalogram, a recurrence plot, and a rendered waveform trace. The selected representations are stacked as channels of a single multichannel input, together with the raw waveform, and processed by a shared asymmetric-attention architecture. Experiments on a $58$-subject stress dataset show that representation fusion improves over the raw waveform. The best configuration, which combines five representations while excluding the unwrapped phase spectrogram, reaches $70.97\%$ test accuracy, compared with $67.36\%$ for the raw waveform. The single PSD spectrogram achieves $69.44\%$, remaining close to the best-fused configuration at a lower computational cost.
The results show that alternative visual forms of the same EDA signal can provide useful inductive biases for stress detection, and that a compact selection of complementary representations can be more effective than the raw waveform alone.

\end{abstract}



\section{Introduction}
Stress is a coordinated physiological and psychological response to perceived demands. It engages the autonomic nervous system and activates neuroendocrine pathways whose intensity and duration depend on the stressor and the individual response \cite{goldstein_2023, hellhammer_wust_2009}. These responses may appear as brief situational episodes or develop into prolonged chronic states, with different physiological patterns and health consequences. Questionnaire instruments such as the Perceived Stress Scale are widely used to quantify subjective stress in clinical and research settings \cite{cohen_kamarck_1983}. However, retrospective self-report is vulnerable to recall bias and cannot capture fine temporal changes in stress levels \cite{shiffman_stone_2008}. Salivary cortisol provides a validated neuroendocrine marker of stress, but sample collection is intrusive, and cortisol dynamics are too delayed for continuous real-time monitoring \cite{hellhammer_wust_2009}.

Stress has become a major public health problem. A large-scale analysis of nationally representative survey data from $146$ countries reported that stress levels roughly doubled over an $18$-year period, with widening demographic and socioeconomic disparities \cite{canaletti_lun_2026}. In occupational settings, psychosocial work-related exposures account for a measurable share of cardiovascular disease and depression cases across Europe \cite{sultantaib_villeneuve_2022}. Chronic psychological stress has also been associated with immune dysregulation, elevated cardiovascular risk, and depressive disorder, reinforcing the need for reliable monitoring methods beyond occasional subjective assessment \cite{cohen_janicki_2007}.

Continuous stress monitoring is especially important in settings where conventional assessment is difficult. In clinical, occupational, and operational environments, self-report can be delayed, incomplete, or affected by social desirability and demand characteristics. Wearable devices offer a practical physiological alternative, but user compliance, motion artifacts, and deployment variability remain persistent challenges \cite{hosseini_gottumukkala_2026}. Field studies have repeatedly identified data quality and signal integrity as key barriers to generalization outside laboratory protocols \cite{neigel_vargo_2025}. A useful stress-detection pipeline should therefore extract as much information as possible from the signals that wearable devices already collect, under conditions that approximate ambulatory use \cite{giannakakis_grigoriadis_2022}.

Electrodermal activity (EDA), also known as galvanic skin response, is among the most established peripheral signals for automatic stress monitoring. It reflects the activity of eccrine sweat glands, which are controlled by the sympathetic branch of the autonomic nervous system and are not under direct voluntary control \cite{boucsein_2012, dawson_schell_2016}. Functional neuroimaging studies have linked EDA to cortical and subcortical regions including the amygdala, insula, and anterior cingulate cortex, placing the signal within the broader neural circuitry involved in stress processing \cite{critchley_2002}. Most EDA-based systems still process the signal as a raw or minimally filtered one-dimensional waveform. Time-frequency and nonlinear-dynamics representations remain less explored, although they may expose structure that is difficult to learn from the waveform alone \cite{sanchez_reolid_2020, singh_kumar_2025}.
Recent work in affective computing has also emphasized the importance of how behavioral and physiological information is organized before classification. Examples include graph-based facial representations for stress recognition, modality-agnostic physiological fusion for cognitive workload assessment, and lightweight modeling of pain-related brain activity~\cite{kassiotis_stressgat_acii_2026, gkikas_workload_acii_2026, gkikas_arzate_eeite_pain_2026}.

This study examines representation choice and fusion for EDA-based stress detection. Six image-based representations are derived from each EDA recording, covering time-frequency and nonlinear-dynamics views of the same signal. These representations are evaluated individually and in channel-stacked combinations with the raw waveform, using a shared asymmetric-attention architecture without separate encoder branches or late fusion. The aim is to assess what is gained by moving beyond the raw waveform, and whether several complementary representations can improve stress recognition when treated as channels of a single multichannel input.

\section{Related Work}
\label{related_work}

Electrodermal activity is widely used in affective computing because it is sensitive to sympathetic arousal and can be recorded continuously with wearable sensors \cite{cowley_torniainen_2016}. In automatic emotion recognition, EDA is commonly used to estimate arousal and valence from tonic and phasic features of the signal. A recent systematic review and meta-analysis reported that EDA-based models perform more reliably for arousal than for valence, in line with the signal's physiological role as an autonomic arousal marker rather than a direct index of affective tone \cite{damelio_galan_2025}. EDA is also used in automatic pain assessment, a related affective-computing task, where recent models have fused EDA with functional near-infrared spectroscopy through gradient-infused attention \cite{khan_chetty_2026} and with electrocardiogram signals through cross-modal transformer fusion \cite{farmani_bargshady_2025}. In these cases, EDA is usually treated as one physiological channel among several, rather than as a signal whose internal representations are studied in detail.

Pain and stress can elicit overlapping affective and autonomic responses. Accordingly, automatic pain assessment has been studied using a wide range of behavioral and physiological signals~\cite{gkikas_tsiknakis_slr_2023}. Electrocardiography (ECG) has been combined with demographic information in both single-task and multi-task formulations~\cite{gkikas_chatzaki_2022,gkikas_chatzaki_2023}, while facial video has been fused with heart-rate signals~\cite{gkikas_tachos_2024}. Other work has combined facial video with Functional near-infrared spectroscopy (fNIRS) \cite{gkikas_tsiknakis_painvit_2024}, and a unified tokenization framework has recently been used to process facial videos, raw fNIRS waveforms, and fNIRS spectrograms within the same architecture~\cite{gkikas_arzate_pain_icmi_2026}. fNIRS-specific models have also explored haemoglobin-difference representations for pain recognition~\cite{bargshady_aziz_2025}.

Stress detection is one of the main application areas for EDA \cite{sanchez_reolid_2020}. Recent work has increasingly applied deep learning directly to wearable-device EDA data, using either handcrafted descriptors or the complete raw signal in custom architectures for early stress detection. In some cases, EDA alone has outperformed multimodal combinations with electrocardiogram and photoplethysmography signals \cite{singh_kumar_2025}. A smaller group of studies has converted EDA into image-based representations. Short-time Fourier transform and Mel spectrograms of the phasic EDA component, combined with convolutional networks such as VGG16, have improved emotion-classification accuracy over feature-based baselines \cite{ganapathy_veeranki_2020}. Signal-to-image encodings have also been explored for stress detection, with Gramian Angular Field, Markov Transition Field, and recurrence plot representations compared on the WESAD dataset; Gramian Angular Field was the strongest encoding, although the analysis used multivariate physiological windows rather than EDA alone \cite{serenelli_quadrini_2024}. Outside the EDA and stress literature, fusing several signal-to-image transformations by stacking them as channels has been shown to outperform individual transformations and to extend naturally to multivariate and multi-sensor time-series classification \cite{mariani_appiah_2025}.

A related strategy was explored in \cite{gkikas_kyprakis_eda_2025}, where multiple EDA representations were generated and composited into a single multi-representation diagram for automatic pain recognition. This approach was competitive with and, in several cases, superior to conventional multimodal fusion. Image-domain physiological processing has also been used at larger scale: a vision foundation model trained across behavioral and physiological modalities, including EDA, electrocardiogram, electromyogram, and functional near-infrared spectroscopy, maps each modality into a shared image-based representation space for pain assessment \cite{gkikas_rojas_painformer_2025}; similarly, a lightweight embedding model for biosignal analysis has been pretrained on $4.4$ million biosignal images spanning several modalities and tasks \cite{gkikas_tiny_2025}. 
Representation design has also been studied directly in the visual domain, including synthetic thermal and RGB videos for pain assessment~\cite{gkikas_tsiknakis_thermal_2024} and reorganized facial spatiotemporal representations intended to preserve more useful structure for recognition~\cite{gkikas_reface_acii_2026}.
These studies either spatially composite representations, stack signal images outside the EDA stress-detection setting, or learn a common embedding space across modality-specific images. 
The same question has also been explored in the temporal domain. Multi-window respiration models have combined short and long signal segments for pain recognition~\cite{gkikas_kyprakis_resp_2025}, while multi-scale EEG models have used sample-adaptive fusion across several window lengths for emotion recognition~\cite{gkikas_guo_eeg_prai_2026}.
A systematic comparison of the raw EDA waveform, multiple EDA-derived visual representations, and channel-stacked combinations under a common subject-level stress-recognition protocol remains missing.


\section{Methodology}
\label{sec:methodology}

\subsection{EDA Signal Representations}
\label{sec:representations}

\begin{figure}
\begin{center}
\includegraphics[scale=0.33]{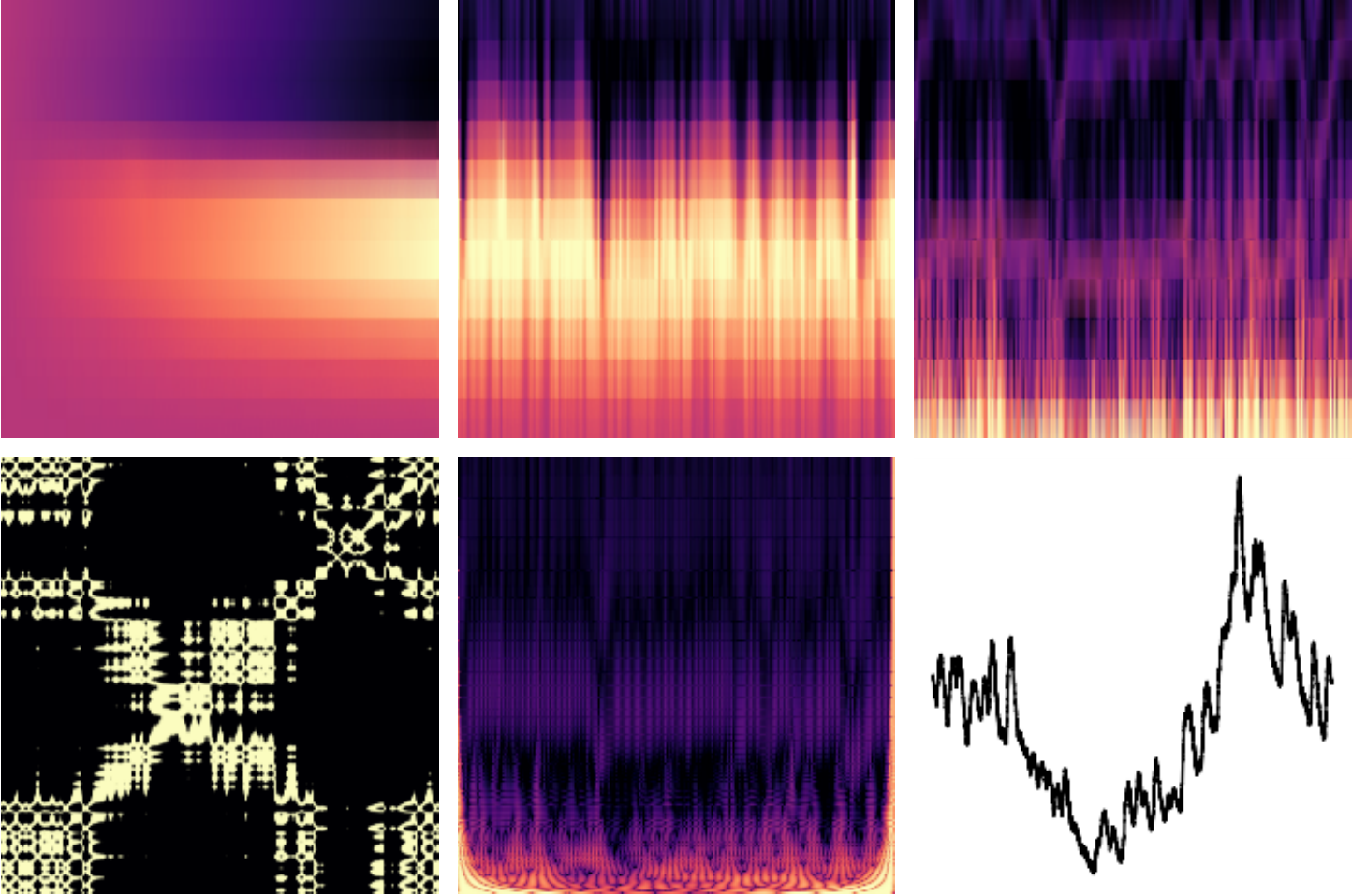}
\end{center}
\caption{Example of the six visual representations derived from a single EDA recording. Top row (left to right): \textit{Angle}, \textit{Phase}, \textit{PSD}. Bottom row (left to right): \textit{Recurrence}, \textit{Scalogram}, \textit{Wave}.}
\label{eda}
\end{figure}

Each EDA recording is a single-channel signal sampled at $1$~kHz throughout the full $120$-second task duration, yielding $119{,}988$ raw samples. Seven representations are derived from the raw signal. The first is the raw waveform itself, used both as a standalone 1D input and as one channel in every combined configuration. The other six are 2D image representations computed after downsampling the signal to $f_{ds}=50$~Hz, matching the approximate $0.05$--$5$~Hz bandwidth of EDA.
The six visual representations are defined as follows.
\textit{(1) PSD.} A magnitude spectrogram is computed via short-time Fourier transform (STFT) with a $128$-sample window and $120$-sample overlap, band-limited to $0.05$--$5$~Hz and log-compressed.
\textit{(2) Angle.} The unwrapped phase of the same STFT is computed by applying phase unwrapping along the time axis, removing the wrapping discontinuities that would otherwise appear as a checkerboard artifact on the slowly varying EDA signal.
\textit{(3) Phase.} An instantaneous-frequency map is derived from the time derivative of the unwrapped phase, with time-frequency bins below the $50$th percentile magnitude masked to zero and the result smoothed with a Gaussian filter ($\sigma=1.5$) to suppress noise in low-power regions.
\textit{(4) Scalogram.} A continuous wavelet transform is applied using a Morlet wavelet across $128$ log-spaced scales corresponding to the $0.05$--$5$~Hz band, producing a multi-resolution time-frequency energy map.
\textit{(5) Recurrence.} A recurrence plot is computed from the z-score normalized signal using a point threshold at the $20$th percentile, visualizing recurring states in the signal's dynamics.
\textit{(6) Wave.} The full-resolution waveform is rendered directly as a line plot, converting the 1D trace into a 2D image without any time-frequency or nonlinear-dynamics transformation.
All six image representations are rendered at $224\times224$ resolution using the same colormap and canvas-capture pipeline, with axes stripped before saving as lossless images. \textit{PSD}, \textit{Phase}, and \textit{Scalogram} are additionally contrast-normalized to the $5$th--$95$th percentile range, while \textit{Angle} is rendered without percentile clipping. The shared rendering pipeline mitigates differences due to image style and keeps the comparison focused on the underlying signal transformation. Figure \ref{eda} shows one example of each representation computed from the same recording.

\subsection{Multi-Representation Fusion}
\label{sec:fusion}

Representations are combined by channel stacking rather than by spatial concatenation or separate encoder branches. Each selected image representation is flattened from its native $224\times224\times3$ shape into a one-dimensional vector of length $150{,}528$. It is then stacked as an additional channel alongside the raw waveform, which is zero-padded from its native length of $119{,}988$ samples to the common length $L=150{,}528$ whenever at least one image representation is included. A configuration with $R$ image representations produces a tensor $\mathbf{X}\in\mathbb{R}^{B\times C\times L}$ with $C=1+R$ channels. This tensor is processed as a 1D input by the same architecture used for the raw waveform alone, analogous to a multichannel physiological signal such as EEG, where recording channels are stacked along $C$. Channel-wise layer normalization is applied before the backbone, so that raw EDA amplitudes and flattened pixel-intensity channels are normalized independently instead of being forced to share a common statistic. When a single image representation is evaluated without the waveform, the native $224\times224\times3$ image is processed directly as a 2D input, following Section~\ref{sec:asymmetric}. In the rest of the paper, waveform $\oplus$ \textit{Angle} denotes the channel-stacked combination of the two representations. The symbol $\oplus$ is used to distinguish channel stacking from elementwise addition, which is not performed here.

\subsection{Asymmetric Attention}
\label{sec:asymmetric}

For either input type, geometric information is added by encoding each token position with Fourier features using $K=6$ frequency bands and a maximum frequency $f_{\max}=10$. The token sequence, whether the $N=H\times W=50{,}176$ tokens of a single $224\times224$ image or the $L$ positions of the raw waveform or a channel-stacked combination, is partitioned into $S=32$ contiguous segments. These segments divide the input along the token axis and do not correspond to external temporal or spatial windows of the original signal.

The same single-layer backbone is used for every configuration: the raw waveform, each individual image-based representation, and every channel-stacked combination. It consists of one cross-attention block followed by $R=8$ self-attention blocks. A single latent state is associated with each segment and is instantiated at runtime by replicating a shared initialization vector derived from a set of $M_0=64$ learnable global parameters
$\{\boldsymbol{\ell}_m\}_{m=1}^{M_0}$:
\begin{equation}
\boldsymbol{\ell}_{\mathrm{init}} = \frac{1}{M_0}\sum_{m=1}^{M_0}
\boldsymbol{\ell}_m \in \mathbb{R}^{d_0},
\end{equation}
where $d_0=128$. These segment states are not independently learnable; segment-specific representations emerge through the attention updates.

\subsubsection{Cross-attention.} Each segment state aggregates information exclusively from its corresponding token subset through cross-attention:
\begin{equation}
\mathbf{e}_s = \mathbf{e}_s^{(0)} +
\mathrm{Attn}\bigl(\mathbf{e}_s^{(0)},\ \tilde{\mathbf{T}}_s\bigr),
\end{equation}
where $\mathbf{e}_s^{(0)}\in\mathbb{R}^{B\times 1\times d_0}$
provides the queries and $\tilde{\mathbf{T}}_s\in\mathbb{R}^{B\times
n_s\times C'}$ provides the keys and values. This operation is asymmetric: the query side consists of a single vector of dimension $d_0$, while the key-value side spans $n_s \gg 1$ token vectors of dimension $C'$. The resulting attention matrix is $1\times n_s$, not square, and the query and key-value spaces differ in both size and dimensionality. All $S=32$ segments are processed in parallel by packing into the batch dimension, without altering the underlying computation. Cross-attention uses $8$ heads, each of dimension $16$.

\subsubsection{Self-attention.} After cross-attention, all segment states are stacked to form the segment-state matrix $\mathbf{E} \in \mathbb{R}^{B\times S\times d_0}$. Self-attention is applied across all $S$ segment states, enabling global information exchange:
\begin{equation}
\mathbf{E} \leftarrow \mathbf{E} +
\mathrm{Attn}\bigl(\mathbf{E},\ \mathbf{E}\bigr),
\end{equation}
repeated $R=8$ times. Self-attention uses $8$ heads, each of dimension $16$.

\begin{figure*}
\begin{center}
\includegraphics[scale=0.60]{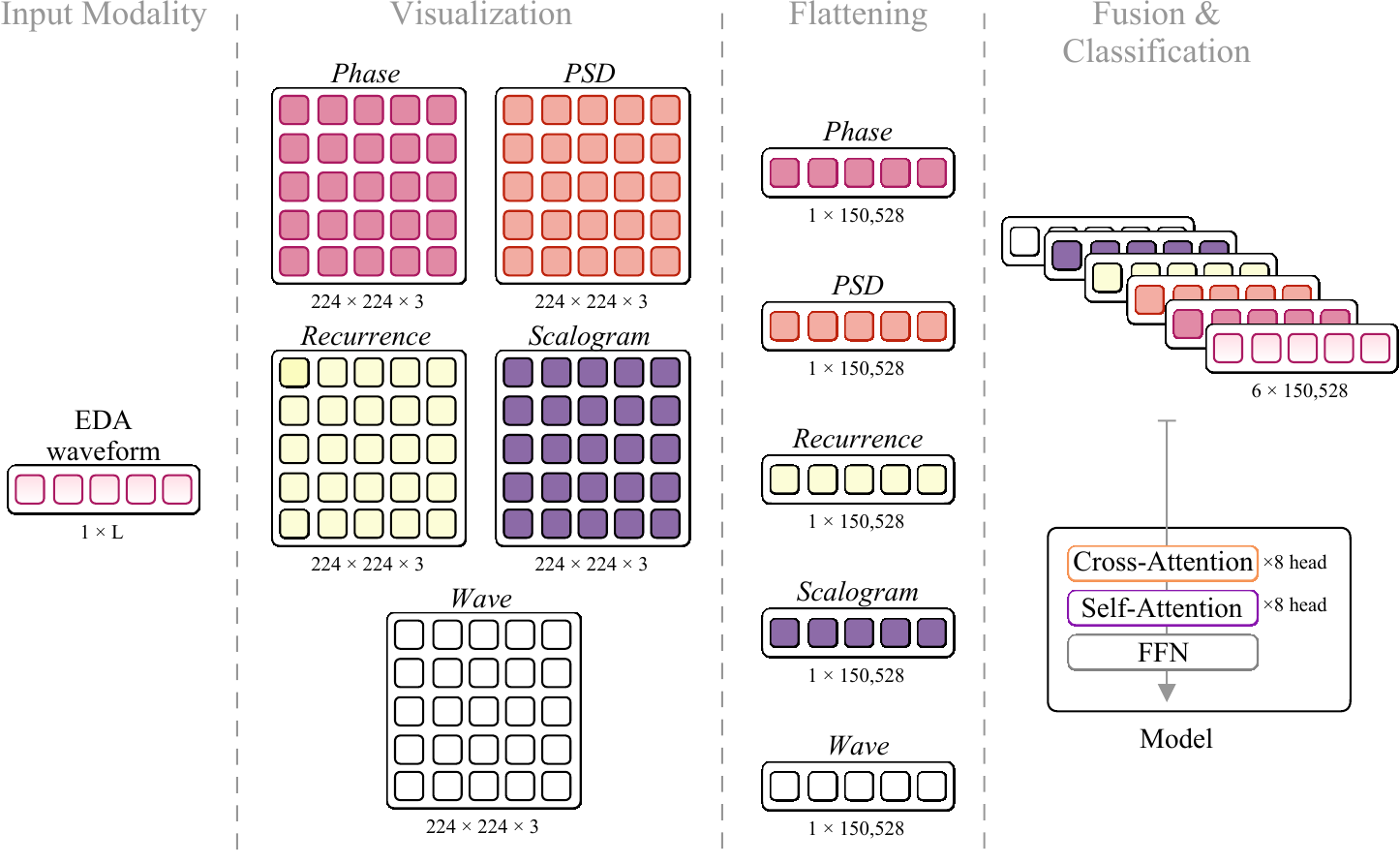}
\end{center}
\caption{Overview of the proposed pipeline. Each EDA recording produces the raw waveform and, in parallel, six visual representations. Representations are flattened into one-dimensional channels, stacked together with the raw waveform, and processed by the shared asymmetric-attention backbone for binary stress classification.}
\label{overview}
\end{figure*}

After the self-attention blocks, the final segment states $\mathbf{E} \in \mathbb{R}^{B\times S\times 128}$ are averaged across segments and passed through a linear classification head. Both attention operations use pre-layer normalization and residual connections, with attention and feedforward dropout of $0.10$ applied uniformly. The complete set of architectural hyperparameters, shared by every configuration evaluated in this study, is reported in Table~\ref{tab:architecture}, and the full pipeline is illustrated in Fig.~\ref{overview}.

\begin{table}
\caption{Architectural hyperparameters, shared by every configuration evaluated in this study.}
\label{tab:architecture}
\begin{center}
\begin{threeparttable}
\begin{tabular}{>{\raggedright\arraybackslash}m{5.5cm} P{2.0cm}}
\toprule
\textbf{Hyperparameter} & \textbf{Value} \\
\midrule
\midrule
Depth                                  & 1     \\\hdashline
Number of latents                      & 64    \\\hdashline
Latent dimension ($d_0$)               & 128   \\\hdashline
Cross-attention heads                  & 8     \\\hdashline
Cross-attention head dimension         & 16    \\\hdashline
Self-attention heads                   & 8     \\\hdashline
Self-attention head dimension          & 16    \\\hdashline
Self-attention blocks per cross ($R$)  & 8     \\\hdashline
Segments ($S$)                         & 32    \\\hdashline
Attention dropout                      & 0.10  \\\hdashline
Feedforward dropout                    & 0.10  \\\hdashline
Fourier frequency bands ($K$)          & 6     \\\hdashline
Maximum frequency ($f_{\max}$)         & 10    \\
\bottomrule
\end{tabular}
\begin{tablenotes}[para,flushleft]
\scriptsize
\item This single-layer backbone is identical across every representation and combination; differences in Params (M) and GFLOPs across rows in Table~\ref{table:eda_results} come entirely from differences in input shape (1D waveform length vs.\ 2D image size vs.\ number of stacked channels), not from any change in architecture.
\end{tablenotes}
\end{threeparttable}
\end{center}
\end{table}

\subsection{Augmentation, Regularization, and Training}
\label{sec:augmentation}

The waveform and image representations have different tensor shapes, so each uses an augmentation pipeline matched to its native format. In combined configurations, the waveform channel is augmented with the 1D pipeline, while each image channel is augmented with the 2D pipeline before flattening and channel stacking. This preserves the natural domain of each transformation during augmentation. Label smoothing, attention dropout, feed-forward dropout, optimizer, learning-rate schedule, and batch size are shared across all configurations. The complete setup is reported in Table~\ref{table:augm_regul}.

\begin{table}
\caption{Augmentation, regularization, and training configuration.}
\label{table:augm_regul}
\centering
\begin{threeparttable}
\begin{tabular}{P{3.4cm} P{4.0cm}}
\toprule
Method / Parameter & Value \\
\midrule
\midrule
\multicolumn{2}{l}{\textit{Waveform (1D) augmentations}} \\
\midrule
AddNoise            & $p \in [0.10, 0.50]$, SNR factor $\in [1, 1000]$ \\\hdashline
Masking             & $p \in [0.10, 0.50]$, mask $15$--$30\%$ of length \\
\midrule
\multicolumn{2}{l}{\textit{Image (2D) augmentations}} \\
\midrule
AugMix              & $p = 0.10$, magnitude $3$ \\\hdashline
TrivialAugmentWide  & $p = 0.10$, $31$ magnitude bins \\\hdashline
CenterCrop          & $p = 0.10$, crop $200\times200$, resize $224\times224$ \\\hdashline
Noise               & $p = 0.10$, $\sigma = 100$ \\\hdashline
Masking             & $p = 0.10$, $20$ blocks of $32\times32$ \\
\midrule
\multicolumn{2}{l}{\textit{Regularization}} \\
\midrule
Label Smoothing     & $0.10$ \\\hdashline
Att-Dropout         & $0.10$ \\\hdashline
FF-Dropout          & $0.10$ \\
\midrule
\multicolumn{2}{l}{\textit{Training}} \\
\midrule
Optimizer           & AdamW \\\hdashline
Learning rate       & $1\mathrm{e}{-4}$ \\\hdashline
LR schedule         & cosine \\\hdashline
Weight decay        & $0.05$ \\\hdashline
Warmup epochs       & $20$ \\\hdashline
Cooldown epochs     & $10$ \\\hdashline
Batch size          & $32$ \\\hdashline
Epochs              & $200 / 400$ \\
\bottomrule
\end{tabular}
\begin{tablenotes}[para,flushleft]
\scriptsize
\item Waveform augmentations are applied to the $119{,}988$-sample raw signal;
image augmentations are applied to the $224\times224$ PIL image. Both are applied
during training only; validation and testing use no augmentation. In combined configurations, each channel type is augmented
in its native domain before flattening and channel stacking.
Att-Dropout/FF-Dropout: dropout probability in
attention/feed-forward sublayers. $p \in [x_1, x_2]$ denotes
$p \sim \mathcal{U}(x_1, x_2)$ sampled per sample.
\end{tablenotes}
\end{threeparttable}
\end{table}

\section{Experimental Evaluation \& Results}
This section evaluates the choice of representation and fusion for EDA-based stress detection. All experiments use binary classification. Validation performance is reported with macro-averaged accuracy, precision, and F1 score, together with their arithmetic mean, denoted \textit{Average}. Test performance is reported with macro-averaged accuracy and is used as the primary comparison criterion.

\subsection{Dataset and Protocol}
\label{ssec:data_collection}
This study uses a stress dataset of $58$ adults ($24$ men, $34$ women), aged $26.9\pm4.8$ years. The protocol includes four stress-induction phases: social exposure, emotional recall, mental workload, and stressful video stimuli. Each participant completed $11$ tasks: $4$ neutral, $6$ stress-inducing, and $1$ relaxation task, as reported in Table~\ref{tbl:tasks}. Social exposure consisted of a psychologist-led interview focused on negative personality traits. Emotional recall required participants to relive a past stressful event in real time. Mental workload was induced with a modified Stroop Color-Word Test \cite{stroop_1935} and the Paced Auditory Serial Addition Test \cite{tombaugh_2006}. The stressful stimuli phase used videos depicting accidents and acrophobia; a relaxing video was included as a physiological recovery baseline between induction phases and was excluded from binary classification.
Stress induction was verified through Heart Rate monitoring, which showed a statistically significant increase during stress tasks ($p<0.05$). Subjective validation was performed using Self-Assessment Manikin scales, in which participants reported higher arousal and lower valence during stressful phases than during neutral baselines.
Electrodermal activity was recorded continuously from a single channel at $1$~kHz and was the only input modality used here. The task is binary classification between neutral and stress conditions. The study received approval from the local Research Ethics Committee (approval no.~155/12-09-2022), and all participants provided informed consent. The dataset is available for non-commercial research upon request.\footnote{\url{https://github.com/ggian/stress_dataset}}
Subjects are split at the subject level, with no participant appearing in more than one set. To reduce performance inflation due to subject-difficulty imbalance, a stratified split protocol is used: subjects are ranked by the combined z-score from leave-one-subject-out difficulty estimation and assigned to four quartiles. The final partition contains $38$ training, $8$ validation, and $12$ testing subjects, with all sets containing subjects from all four difficulty groups in proportion. Table~\ref{tab:subject_split_stress} reports the exact split. The resulting data contain $414$ training samples ($188$ neutral, $226$ stress), $83$ validation samples ($37$ neutral, $46$ stress), and $132$ testing samples ($60$ neutral, $72$ stress).

\begin{table}
\caption{Experimental tasks employed in this study.}
\label{tbl:tasks}
\begin{center}
\begin{threeparttable}
\begin{tabular}{P{0.5cm} P{3.5cm} P{2.0cm} P{1.0cm}}
\toprule
\# & Task & Duration (sec) & State \\
\midrule
\midrule
\multicolumn{4}{l}{\textit{Social Exposure}} \\
1  & Neutral reference        & 120 & N \\\hdashline
2  & Baseline description     & 120 & N \\\hdashline
3  & Interview                & 120 & S \\
\midrule
\multicolumn{4}{l}{\textit{Emotional Recall}} \\
4  & Neutral reference        & 120 & N \\\hdashline
5  & Recall stressful event   & 120 & S \\
\midrule
\multicolumn{4}{l}{\textit{Mental Workload}} \\
6  & Reading reference        & 120 & N \\\hdashline
7  & Stroop Color-Word Test   & 120 & S \\\hdashline
8  & PASAT task               & 120 & S \\
\midrule
\multicolumn{4}{l}{\textit{Stressful Stimuli}} \\
9  & Relaxing video$^{*}$     & 120 & R \\\hdashline
10 & Adventure video          & 120 & S \\\hdashline
11 & Psychological pressure   & 120 & S \\
\bottomrule
\end{tabular}
\begin{tablenotes}[para,flushleft]
\scriptsize
\item N\,=\,neutral\quad S\,=\,stress \quad R\,=\,relaxed. *: Used as a physiological recovery baseline between induction
phases; excluded from binary classification.
\end{tablenotes}
\end{threeparttable}
\end{center}
\end{table}

\begin{table*}
\caption{Subject-level split by difficulty group. Subjects are ranked by
combined z-score and assigned to four quartile-based groups
(Q1\,=\,hardest, Q4\,=\,easiest).}
\label{tab:subject_split_stress}
\begin{center}
\begin{threeparttable}
\begin{tabular}{P{1.65cm} P{3.6cm} P{3.6cm} P{3.6cm} P{3.6cm}}
\toprule
\multirow[c]{3}{*}{Split} &
\multicolumn{4}{c}{Difficulty Group} \\
\cmidrule(lr){2-5}
 & Q1 -- Hard & Q2 -- Med-Hard & Q3 -- Med-Easy & Q4 -- Easy \\
\midrule
\midrule
Training (38) &
P017, P018, P022, P026, P034, P035, P042, P045, P050, P056 &
P001, P002, P003, P007, P012, P021, P033, P040, P048 &
P004, P014, P016, P032, P036, P046, P047, P052, P053, P054 &
P005, P010, P020, P028, P029, P037, P039, P041, P057 \\\hdashline
Validation (8) &
P038, P055 &
P009, P023 &
P006, P013 &
P019, P030 \\\hdashline
Testing (12) &
P008, P025, P044 &
P011, P024, P043 &
P015, P031, P058 &
P027, P051, P059 \\
\bottomrule
\end{tabular}
\begin{tablenotes}[para,flushleft]
\scriptsize
\item Q1: $z < -0.46$;\quad Q2: $-0.46 \leq z < -0.05$;\quad
Q3: $-0.05 \leq z < +0.40$;\quad Q4: $z \geq +0.40$.
\end{tablenotes}
\end{threeparttable}
\end{center}
\end{table*}

\subsection{Representation Comparison}
\label{sec:representation_comparison}
Table~\ref{table:eda_results} reports performance and computational cost for every single representation and every combined configuration after $400$ training epochs. The raw waveform reaches $67.36\%$ test accuracy. Seven of the eight combined configurations exceed this value; the only exception is waveform $\oplus$ \textit{Phase}, which drops to $64.72\%$. The best configuration combines the waveform with \textit{PSD}, \textit{Recurrence}, \textit{Scalogram}, and \textit{Wave}, while excluding \textit{Angle}. It reaches $70.97\%$, improving the raw waveform by $3.61$ points. Fusing all six representations also improves over the raw waveform, reaching $68.89\%$, but remains below the excluded-\textit{Angle} configuration.

Among individual representations, \textit{PSD} is the strongest on test accuracy, reaching $69.44\%$. It remains within $1.53$ points of the best fused configuration while requiring only one representation. \textit{Wave} and waveform $\oplus$ \textit{Wave} both reach $68.89\%$, showing that a rendered image of the waveform trace is competitive even though it contains no new signal source. \textit{Angle} is the weakest single representation, with $57.78\%$ test accuracy, roughly $10$ points below the other single inputs. This result motivates its exclusion from the best-performing fusion.

All results above use the $400$-epoch training budget. Table~\ref{table:eda_epochs} reports test accuracy at $200$ and $400$ epochs for each configuration. The longer budget was introduced after preliminary experiments showed that some representations, especially \textit{Wave}, continued to improve after $200$ epochs.

\begin{table*}
\caption{Performance and computational cost using the EDA modality (400 epochs).}
\label{table:eda_results}

\begin{center}
\begin{threeparttable}
\begin{tabular}{P{3.2cm} P{1.5cm} P{0.9cm} P{1.0cm} P{1.0cm} P{0.7cm} P{1.0cm} P{1.2cm}}
\toprule

\multirow[c]{3}{*}{Representation} &
\multicolumn{2}{c}{Computational Cost} &
\multicolumn{4}{c}{Validation} &
\multicolumn{1}{c}{Testing} \\

\cmidrule(lr){2-3}\cmidrule(lr){4-7}\cmidrule(lr){8-8}
& Params (M) & GFLOPs & Accuracy & Precision & F1 & \textit{Average} & Accuracy \\

\midrule
\midrule
Waveform             &2.00 &0.97 &61.78 &61.66 &61.40 &\textit{61.61} &67.36 \\\hdashline
Angle                &1.86 &0.86 &66.13 &66.00 &66.02 &\textit{66.05} &57.78 \\\hdashline
Phase                &1.86 &0.86 &71.30 &72.21 &71.48 &\textit{71.66} &62.78 \\\hdashline
PSD                  &1.86 &0.86 &68.33 &70.26 &68.39 &\textit{68.99} &69.44 \\\hdashline
Recurrence           &1.86 &0.86 &66.42 &67.13 &66.52 &\textit{66.69} &65.69 \\\hdashline
Scalogram            &1.86 &0.86 &65.63 &67.95 &65.46 &\textit{66.35} &64.86 \\\hdashline
Wave                 &1.86 &0.86 &66.42 &67.13 &66.52 &\textit{66.69} &68.89 \\
\midrule
Waveform $\oplus$ Angle       &2.06 &1.27 &62.31 &62.38 &61.44 &\textit{62.04} &67.64 \\\hdashline
Waveform $\oplus$ Phase       &2.06 &1.27 &63.40 &63.40 &62.65 &\textit{63.15} &64.72 \\\hdashline
Waveform $\oplus$ PSD         &2.06 &1.27 &66.42 &67.13 &66.52 &\textit{66.69} &67.92 \\\hdashline
Waveform $\oplus$ Recurrence  &2.06 &1.27 &60.75 &62.35 &60.35 &\textit{61.15} &68.47 \\\hdashline
Waveform $\oplus$ Scalogram   &2.06 &1.27 &67.51 &68.49 &67.62 &\textit{67.87} &67.92 \\\hdashline
Waveform $\oplus$ Wave        &2.06 &1.27 &66.42 &67.13 &66.52 &\textit{66.69} &68.89 \\\hdashline
All                    &2.06 &1.66 &68.60 &69.91 &68.72 &\textit{69.08} &68.89 \\\hdashline
All (no angle)         &2.06 &1.58 &65.31 &65.13 &64.98 &\textit{65.14} &\textbf{70.97}
\\

\bottomrule
\end{tabular}
\begin{tablenotes}[para,flushleft]
\scriptsize
\item \textit{Average}: arithmetic mean of validation Accuracy, Precision, and F1. The bold value indicates the highest overall test accuracy.
\item Params (M) and GFLOPs for the visual representations are identical to \textit{Angle}: all six are $224\times224\times3$ images processed by the identical backbone (Table~\ref{tab:architecture}). \textit{Waveform} uses the same backbone throughout; its different Params/GFLOPs reflect its different input shape (a 1D sequence rather than a 2D image), not a different architecture.
\end{tablenotes}
\end{threeparttable}
\end{center}
\end{table*}

\begin{table}
\caption{Test accuracy at 200 vs.\ 400 training epochs.}
\label{table:eda_epochs}
\begin{center}
\begin{threeparttable}
\begin{tabular}{P{3.0cm} P{1.1cm} P{1.1cm} P{0.9cm}}
\toprule
Representation & 200 ep. & 400 ep. & $\Delta$ \\
\midrule
\midrule
Waveform              &67.08 &67.36 &$+0.28$  \\\hdashline
Angle                 &57.78 &57.78 &$+0.00$  \\\hdashline
Phase                 &61.53 &62.78 &$+1.25$  \\\hdashline
PSD                   &69.44 &69.44 &$+0.00$  \\\hdashline
Recurrence            &63.75 &65.69 &$+1.94$  \\\hdashline
Scalogram             &63.75 &64.86 &$+1.11$  \\\hdashline
Wave                  &58.89 &68.89 &$+10.00$ \\
\midrule
Waveform $\oplus$ Angle        &69.03 &67.64 &$-1.39$  \\\hdashline
Waveform $\oplus$ Phase        &63.47 &64.72 &$+1.25$  \\\hdashline
Waveform $\oplus$ PSD          &66.67 &67.92 &$+1.25$  \\\hdashline
Waveform $\oplus$ Recurrence   &66.81 &68.47 &$+1.66$  \\\hdashline
Waveform $\oplus$ Scalogram    &64.31 &67.92 &$+3.61$  \\\hdashline
Waveform $\oplus$ Wave         &68.75 &68.89 &$+0.14$  \\\hdashline
All                     &68.19 &68.89 &$+0.70$  \\\hdashline
All (no angle)          &69.44 &70.97 &$+1.53$  \\
\bottomrule
\end{tabular}
\end{threeparttable}
\end{center}
\end{table}

\begin{figure*}
\begin{center}
\includegraphics[scale=0.60]{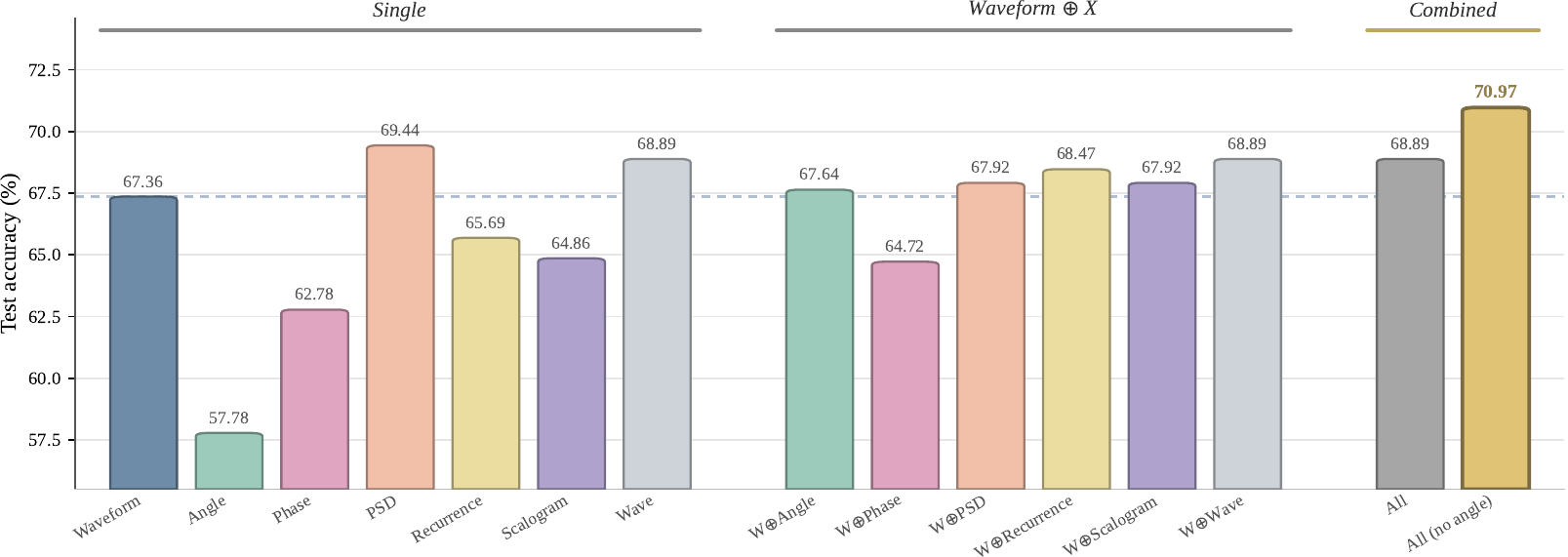}
\end{center}
\caption{Test accuracy across single representations, waveform $\oplus$ representation combinations, and the two full-fusion configurations, at the $400$-epoch training budget. The excluded-\textit{Angle} configuration reaches the highest test accuracy at $70.97\%$; \textit{PSD} alone remains within $1.53$ points at $0.86$ vs.\ $1.58$ GFLOPs.}
\label{performances}
\end{figure*}

\subsection{Overall Analysis \& Discussion}

Channel-stacked fusion gives the best test accuracy in this study, but the margin over the strongest single representation is modest. The excluded-\textit{Angle} configuration achieves $70.97\%$ on the test set, while \textit{PSD} alone achieves $69.44\%$, a difference of $1.53$ points at roughly half the GFLOPs. Fusion of all six representations reaches $68.89\%$, the same as \textit{Wave} alone. Fusion is therefore not automatically the best practical choice. In accuracy-critical settings, the excluded-\textit{Angle} configuration is preferred; in compute-constrained settings, a carefully selected single representation remains competitive.

We note that in this study, since all configurations were evaluated using the same fixed subject-independent split, the results are treated as a benchmark comparison, with held-out test accuracy as the primary metric. Under this protocol, the excluded-\textit{Angle} fusion achieved the highest test accuracy. Its $1.53$-point improvement over \textit{PSD}, however, should be interpreted only within this experimental setting and not as evidence that the same configuration will necessarily remain superior across other datasets, splits, or architectures.

The results also show that fusion is not uniformly beneficial. Waveform $\oplus$ \textit{Phase} reaches $64.72\%$, below the raw waveform at $67.36\%$, and waveform $\oplus$ \textit{Angle} improves the waveform only marginally, from $67.36\%$ to $67.64\%$. The weaker fused configurations are those built around \textit{Phase} and \textit{Angle}, the two STFT-phase-derived representations. In this setting, phase-based encodings do not add useful complementary information when stacked with the raw waveform and can degrade the fused input.

The number of parameters and GFLOPs reported in Table~\ref{table:eda_results} include the classification model only. The visual representations are generated offline using standard signal-processing operations that can run on a conventional CPU and do not require GPU acceleration. They nevertheless introduce additional preprocessing and storage overhead. Since these costs were not measured, the reported efficiency comparison is limited to model complexity and inference computation.

The exclusion of \textit{Angle} from the strongest configuration follows from its empirical behavior. \textit{Angle} has the lowest test accuracy among all single representations, roughly $10$ points below the next weakest single input. It is also present in the two combined configurations, waveform $\oplus$ \textit{Angle} and All, that remain below \textit{All (no angle)}. Removing it therefore addresses a specific failure of this representation rather than imposing a general rule about which signal views should be included.

The \textit{Wave} representation reaches $68.89\%$ on its own and the same value when fused with the waveform. This representation contains the same underlying information as the raw waveform, but in a rendered 2D form. Its performance indicates that the choice of representation affects the inductive bias available to the model, even when the original signal content remains unchanged. In other words, the way the signal is presented to the architecture matters.

Several limitations should be noted. First, the six representations were evaluated with fixed transformation parameters. Tuning the STFT window, wavelet scales, recurrence threshold, or rendering parameters could improve specific representations and alter the observed ranking. Second, the channel order in the stacked input was fixed. The attention-based backbone can, in principle, route information across channels regardless of their positions, but this was not tested directly. Third, the study uses a single dataset and a binary neutral-versus-stress classification setting. The ranking of representations, the benefit of fusion, and the exclusion of \textit{Angle} should therefore be tested on additional stress and affective-computing datasets.

\section{Conclusion}
This paper examined representation choice and fusion for EDA-based stress detection. The raw one-dimensional waveform and six image-based signal representations were evaluated individually and in channel-stacked combinations. All configurations used the same asymmetric-attention backbone, isolating the effect of the input representation from changes in architecture.
Representation fusion improved over the raw waveform in seven of eight combined configurations. The best result was obtained with the excluded-\textit{Angle} configuration, achieving $70.97\%$ test accuracy, a $3.61$-point improvement over the raw waveform at $67.36\%$. The \textit{PSD} spectrogram achieved $69.44\%$ as a single representation, remaining close to the best fused configuration while incurring lower computational cost. The unwrapped phase spectrogram, \textit{Angle}, was the only representation that consistently weakened fusion and was excluded from the best-performing combination based on its test behavior.
Future work should examine per-representation parameter tuning, longer or representation-specific training schedules, alternative channel-ordering strategies, and evaluation on additional stress datasets. The results provide a reference point for studying biosignal representations in stress detection and show that the choice of representation can be as important as the model architecture itself.

\section*{Acknowledgments}
The authors used large language model (LLM)-based tools for language editing and improvement. All scientific content, results, and conclusions are solely the work of the authors.


\bibliographystyle{IEEEtran}
\bibliography{library}

\end{document}